\documentclass
[aps,prb,twocolumn,floatfix,english,showpacs,10pt,superscriptaddress]{revtex4-2}
\usepackage{graphicx}
\usepackage{booktabs}
\usepackage{amsmath}
\usepackage{physics}
\usepackage{amssymb}
\usepackage{colordvi}
\usepackage{verbatim}
\usepackage{xcolor}
\usepackage{mathrsfs}
\usepackage{epsfig}
\usepackage{lipsum}
\usepackage{amsfonts}
\usepackage{makecell}
\usepackage[unicode=true, breaklinks=false, pdfborder={0 0 1}, backref=false,
colorlinks=true, linkcolor=blue, urlcolor=blue, citecolor=blue]{hyperref}%
\providecommand{\U}[1]{\protect\rule{.1in}{.1in}}
\setcitestyle{numbers,square}
\begin{document}

%\title{Edge and Corner Accumulation of Magnon Precisely Characterized by Double Topological Winding Numbers}
%DMK: My suggestion for title: should we emphasize non-hermitian in the title?
%\title{Topological Winding Tuple to Characterize Boundary Accumulation of Magnons}
\title{Edge and corner skin effects of chirally coupled magnons characterized by a topological winding tuple}
\author{Chengyuan Cai}
\affiliation{School of Physics, Huazhong University of Science and Technology, Wuhan 430074, China}
\author{Dante M. Kennes}
\affiliation{Institut f\"ur Theorie der Statistischen Physik, RWTH Aachen University and JARA-Fundamentals of Future Information Technology, 52056 Aachen, Germany}
\affiliation{Max Planck Institute for the Structure and Dynamics of Matter, Luruper Chaussee 149, 22761 Hamburg, Germany}
\author{Michael A. Sentef}
\affiliation{Institute for Theoretical Physics and Bremen Center for Computational Materials Science,
University of Bremen, 28359 Bremen, Germany}
\affiliation{Max Planck Institute for the Structure and Dynamics of Matter, Luruper Chaussee 149, 22761 Hamburg, Germany}
\author{Tao Yu}
\email{taoyuphy@hust.edu.cn}
\affiliation{School of Physics, Huazhong University of Science and Technology, Wuhan 430074, China}
\date{\today }

\begin{abstract}
We investigate a long-ranged coupled and non-Hermitian two-dimensional array of nanomagnets, fabricated on a thin magnetic substrate and subjected to an in-plane magnetic field. We predict topology-driven edge and corner skin effects of magnetic eigenmodes with the localization position at boundaries precisely characterized by a topological winding tuple $({\cal W}_1,{\cal W}_2)$. By varying the direction of the in-plane field, all magnon states pile up either at different edges of the array with $({\cal W}_1=\pm 1,{\cal W}_2=0)$ or $({\cal W}_1=0,{\cal W}_2=\pm 1)$, or at different corners characterized by $({\cal W}_1=\pm 1,{\cal W}_2=\pm 1)$. Exploiting the non-Hermitian topology is potentially helpful for designing useful magnonic metasurface in the future. 

%The uncovered winding tuple establishes bulk-boundary correspondence for two-dimensional non-Hermitian systems. 

 %That does not belong to an abstract. Maybe it is something for the intro or potential outlook.
 %Our findings deepen the understanding of the non-Hermitian topology and are potentially helpful for  designing useful magnonic metasurface in the future. 
\end{abstract}
\maketitle

\textit{Introduction}.---The discovery of the one-dimensional non-Hermitian skin effect, yielding a localization of a macroscopic number of bulk eigenstates at the edge~\cite{Bergholtz,XZhang_review,Okuma_review,RLin,KDing,Yu_review}, stimulated the recent explorations of open systems, achieving useful functionalities such as funneling of light~\cite{Weidemann}, unidirectional amplification~\cite{XWen,McDonald}, non-local response~\cite{Helbig}, and enhanced device sensitivity~\cite{Ghatak,Yu_Zeng,Budich,HYuan}. The winding number of the frequency spectrum $\omega(\kappa)$, defined for periodic boundary conditions, was found to form a loop in the presence of a skin effect when the wave number $\kappa$ evolves by one period. In one dimension, this winding number characterizes the skin effect's topological origin and precisely determines on which edge the eigenstates localize~\cite{FKK,Shunyu,YXiong,Borgnia,KZhang1,Okuma,CYin,HHu,Yokomizo,HShen,ZGong,Kawabata_2}. Extending the non-Hermitian skin effect from one to higher dimensions yields rich and diverse manifestations of skin modes including edge, corner, surface, or hinge localization~\cite{CHLee1,YFu,Denner,Kawabata,Okugawa,KZhang2,WZhu,YLi,Schindler,BXie,Flebus_van_der_Waals,TLi}, which have been experimentally observed in acoustics~\cite{XZhang} and topoelectrical circuits~\cite{CShang,DZou}, but not yet in magnonics ~\cite{Yu_Zeng,Flebus_van_der_Waals,Flebus_review,Yu_review}. Magnonic systems exploit magnetic excitations, i.e., magnons, as potential low-energy-consumption information carriers~\cite{Lenk,Chumak,Grundler,Demidov,Brataas,Barman}.

The rapid progress in the field also raised theoretical challenges and urgent issues in the topological characterization of the different skin modes~\cite{KZhang2,Haiping,Kawabata,HYWang}. In the non-reciprocal two-dimensional non-Hermitian systems, the two winding numbers defined along two normal directions, i.e., a topological winding tuple, may precisely distinguish different edge and corner skin effects, i.e., a precise prediction of the edge or corner on which the modes localize, which is a straightforward generalization of the one-dimensional winding number ~\cite{Kawabata}. However, it may not apply to the reciprocal systems such as the higher-order corner skin effect that originates from specific geometry~\cite{Okugawa} and the geometry-dependent skin effect that depends on the boundaries~\cite{KZhang2}. Kawabata \textit{et al.} showed that the non-zero Wess-Zumino term leads to the presence of (higher-order) corner skin modes in non-Hermitian systems~\cite{Kawabata}. Zhang \textit{et al.} proposed a general theorem to characterize the existence of a non-Hermitian skin effect in higher dimensions in terms of spectra area in the complex plane~\cite{KZhang2}, \textit{viz.} the non-Hermitian skin effect appears when the spectra under periodic boundary conditions cover a finite area.  

\begin{figure}[h!tp]
	\centering
\includegraphics[width=0.48\textwidth]{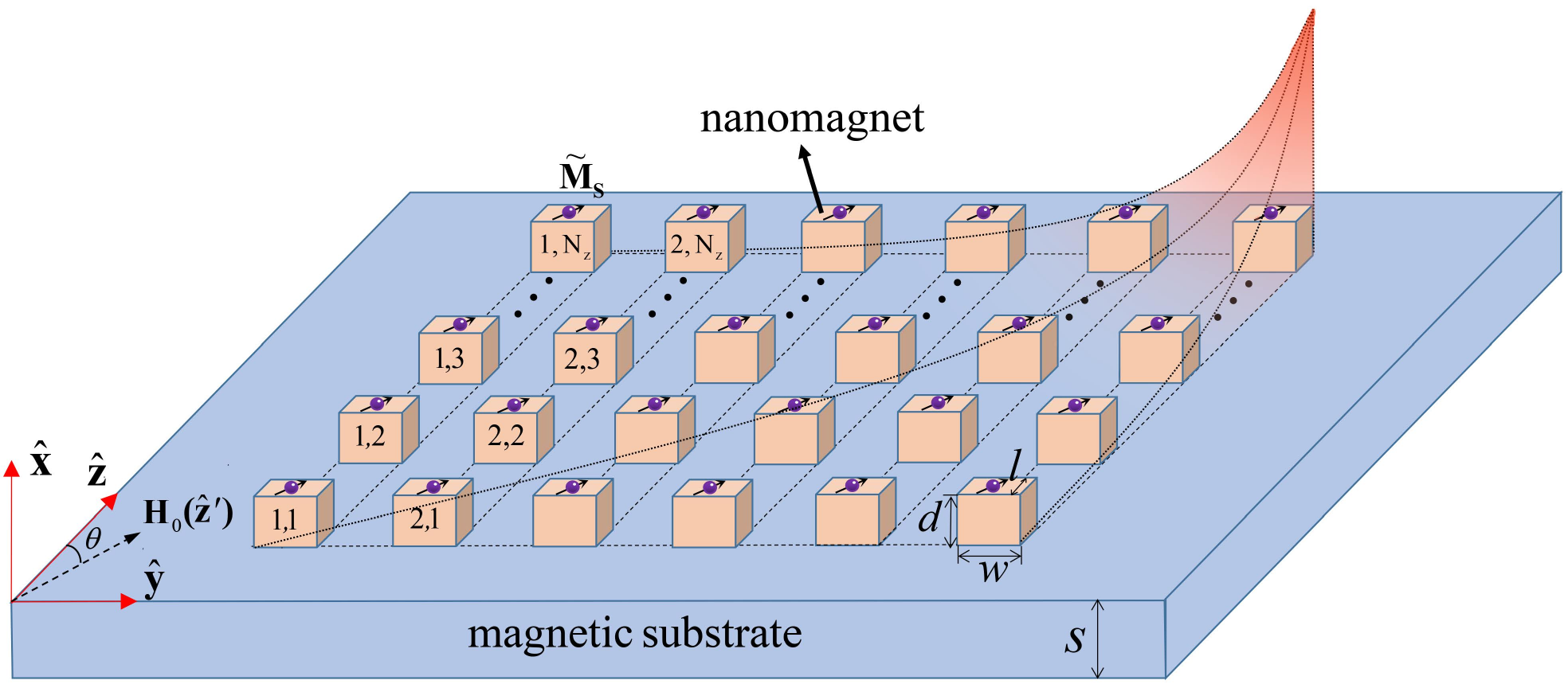}
	\caption{Regularly shaped two-dimensional array of nanomagnets fabricated on a finite area of a magnetic substrate. By varying the direction $\theta$ of the in-plane applied magnetic field ${\bf H}_0$, all magnon eigenstates in the nanomagnets pile up either at the edge or at the corner of the array. The red mode profile implies the localization at one corner. The geometric parameters are given in the text.}
	\label{model}
\end{figure}
In this Letter, we predict different edge or corner skin effects of magnons in ferromagnetic heterostructures composed of a regularly shaped two-dimensional (2D) array of nanomagnets that are fabricated on a thin magnetic substrate and biased by an in-plane magnetic field. The system is illustrated in Fig.~\ref{model}. Mediated by the propagating magnons in the substrate, the indirect interaction between  Kittel magnons~\cite{Kittel} in the nanomagnet is long-range and chiral~\cite{chiral,CPSW}, driving the boundary skin effect. Here the frequency spectrum $\omega(\kappa_1,\kappa_2)$ under periodic boundary conditions is a function of two real wave numbers $\kappa_1$ and $\kappa_2$, which allows us to define a winding tuple $({\cal W}_1,{\cal W}_2)$ by fixing one of the wave numbers. We use such winding tuples to fully characterize different edge and corner aggregations of bulk eigenstates that precisely predict which edge or corner the modes localize on, which can be varied by varying the direction of the in-plane field in our model.  
The winding tuple shows that all of the magnonic bulk eigenstates pile up either at different edges of the array with $({\cal W}_1=\pm 1,{\cal W}_2=0)$ or $({\cal W}_1=0,{\cal W}_2=\pm 1)$, or at different corners characterized by $({\cal W}_1=\pm 1,{\cal W}_2=\pm 1)$.  These predictions can be tested experimentally with conventional metallic nanomagnets on a high-quality thin
magnetic substrate such as yttrium iron garnet (YIG).

\textit{Non-Hermitian magnonic edge and corner eigenstates.}---We consider a finite-sized 2D square array of regular shape composed of $N_y\times N_z$ nanomagnets, e.g. CoFeB, Py, Ni, or Co, of width $w\sim {\cal O}(100)$~nm, length $l\sim {\cal O}(100)$~nm, and thickness $d\sim {\cal O}(10)$~nm, fabricated on the finite area of a magnetic substrate such as YIG thin film of thickness $s\sim {\cal O}(10)$~nm, as illustrated in Fig.~\ref{model}. The distance between neighboring nanomagnet is $\Lambda_y$ and $\Lambda_z$, respectively, in the $\hat{\bf y}$- and $\hat{\bf z}$-directions, and $(a,b)$ indicates the nanomagnet in the $a$-th column and $b$-th row.
An in-plane magnetic field ${\bf H}_0$ with an angle $\theta$ with respect to the $\hat{\bf z}$-direction biases the saturated magnetization ${\bf M}_s$ and $\tilde{\bf M}_s$ of the substrate and nanomagnets. For soft YIG magnetic substrates, ${\bf M}_s$ is parallel to ${\bf H}_0$. $\tilde{M}_s$ is larger than $M_s$ and, due to the shape anisotropy, it has an angle $\tilde{\theta}\ne \theta$ with respect to the $\hat{\bf z}$-direction. We refer to the Supplemental Material (SM)~\cite{supplement} for the calculation of $\tilde{\theta}$.

When $\Lambda_{y,z}\gg \{w,l,d\}$ is of micrometer size, the direct dipolar interaction between the nanomagnets is suppressed to be negligibly small. The nanomagnet then couples dominantly with the magnetic substrate via the dipolar interaction, assuming that the interlayer exchange interaction is suppressed by inserting a thin insulator layer~\cite{JChen}. So the ferromagnetic resonance (FMR) modes or Kittel magnons $\hat{\beta}_{a,b}$~\cite{Kittel} of frequency $\Omega$ in the $(a,b)$-th nanomagnets couple indirectly via the dipolar interaction with the traveling magnons $\hat{m}_{\bf k}$ of wave vector ${\bf k}=(k_y,k_z)$ in the  substrate~\cite{CPSW,dipolar} with the coupling constant given by 
$g_{\bf k}^{(a,b)}=g_{\mathbf{k}}e^{i \left(ak_y \Lambda_y+bk_z \Lambda_z\right)}$, where $g_{\mathbf{k}}$ is real (refer to the SM~\cite{supplement} for detailed derivations). The total Hamiltonian 
\begin{align}
\hat{H}/\hbar&=\sum_{a,b}(\Omega-i\delta_\beta) \hat{\beta}^{\dagger}_{a,b}\hat{\beta}_{a,b}+\sum_{\bf k}(\omega_k-i\delta_m)\hat{m}^{\dagger}_{\bf k}\hat{m}_{\bf k}\nonumber\\
&+\left(\sum_{a,b}\sum_{\bf k}g^{(a,b)}_{\bf k}\hat{m}_{\bf k} \hat{\beta}^{\dagger}_{a,b}+{\rm H.c.}\right)
\end{align}
describes coupled harmonic oscillators, 
where  $\delta_{\beta}=\tilde{\alpha}_G\Omega$ and $\delta_m={\alpha}_G\omega_k$ with the damping constants $\tilde{\alpha}_G$ and $\alpha_G$ for the magnons in the nanomagnet and substrate, and $\omega_k=\mu_0\gamma(H_0+\alpha_{\rm ex}M_s{k}^2)$ is the dispersion of the exchange magnons in the substrate with the vacuum permeability $\mu_0$, the modulus of electron gyromagnetic ratio $\gamma$, and the exchange stiffness $\alpha_{\rm ex}$.

The Kittel magnons in the nanomagnets 
couple effectively via virtually exchanging magnons in the substrate ~\cite{Yu_Zeng,CPSW,dipolar}. 
The effective coupling between magnons in the $(a,b)$-th and $(a',b')$-th nanomagnet is $
\Gamma({\bf r}_{a-a',b-b'})=i\sum_{\bf k}{g_{\bf k}^2e^{i\left[(a-a')k_y \Lambda_y+(b-b')k_z \Lambda_z\right]}}/({\omega-\omega_{k}+i \delta_m})$.
In polar coordinates ${\bf k}=(k,\varphi)$ and ${\bf r}_{a-a',b-b'}=(r_{a-a',b-b'},\phi_{a-a',b-b'})$, performing the contour integral over $k$  with the on-shell approximation $\omega\rightarrow\Omega$ yields~\cite{supplement}
 \begin{align}
 &\Gamma({\bf r}_{a-a',b-b'}=0)=\frac{L_y L_z}{4\pi}\int_{0}^{2\pi} d \varphi \frac{k_\Omega}{v_{k_\Omega}}g^2(k_\Omega,\varphi),\nonumber\\
 &\Gamma({\bf r}_{a-a',b-b'}\ne 0)=\frac{L_y L_z}{2 \pi}\int_{\phi_{a-a',b-b'}-\frac{\pi}{2}}^{\phi_{a-a',b-b'}+\frac{\pi}{2}} d \varphi  \frac{k_\Omega}{v_{k_\Omega}}g^2(k_\Omega,\varphi)\nonumber\\&\times \exp[{i q_\Omega r_{a-a',b-b'} \cos (\varphi-\phi_{a-a',b-b'})}],
 \label{effective_coupling}
 \end{align}
where the lengths of substrate $L_y$ and $L_z$ are along the $\hat{\bf y}$- and $\hat{\bf z}$-directions, $k_\Omega=\sqrt{(\Omega-\mu_0\gamma H_0)/(\mu_0\gamma\alpha_{\rm ex}M_s)}$ is the wave number of the resonant magnon to the FMR frequency $\Omega$ that propagates with group velocity $v_{k_\Omega}=(\partial\omega_k/\partial k)|_{k_\Omega}=2\mu_0\gamma\alpha_{\rm ex}M_sk_\Omega$, and  $q_\Omega=k_\Omega(1+i\alpha_G/2)$.
Therefore, the elements of the effective Hamiltonian matrix of nanomagnet magnons read
 \begin{align}
    &{\cal H}_{\rm eff}\big|_{a=a',b=b'}=\Omega-i \delta_\beta-i \Gamma({\bf r}_{a-a',b-b'}=0),\nonumber\\
    &{\cal H}_{\rm eff}\big|_{a\ne a'~{\rm or}~ b\neq b'}=-i\Gamma\big({\bf r}_{a-a',b-b'}\ne 0\big).
    \label{non_Hermitian_matrix}
	 \end{align}
The substrate, on the one hand, adds an extra dissipation $\Gamma({\bf r}_{a-a',b-b'}=0)$ to the Kittel magnons $\delta_{\beta}$, and, on the other hand, mediates an effective coupling $\Gamma\big({\bf r}_{a-a',b-b'}\ne 0\big)$ between different nanomagnets. The matrix (\ref{non_Hermitian_matrix}) is non-Hermitian such that its diagonalization requires, in general, different left $\eta_{\xi}$ and right $\psi_{\xi}$ eigenvectors, where the state index $\xi=\{1,2,\cdots,N_yN_z\}$. The left and right eigenvectors  obey the biorthonormal condition $\eta_\xi^\dagger\psi_{\xi'}=\delta_{\xi\xi'}$~\cite{Bergholtz,Moiseyev,nonHermrev1}.

 Here we illustrate the results with dimensions used in
experiments Refs.~\cite{XYWei,JChen,HWang} by considering an  array of $30\times30$ CoFeB nanomagnets of thickness $d=30$~nm, width $w=100$~nm, and length $l=200$~nm with the neighboring distance $\Lambda_y=\Lambda_z=2.2$~$\mu$m that are fabricated on the thin YIG film of thickness $s=10$~nm, biased by the in-plane magnetic field $\mu_0H_0=0.05$~T. The saturated magnetization of CoFeB 
 $\mu_0\tilde{M}_s=1.6$~T~\cite{CoFeP_Ms} is much larger than that of YIG $\mu_0{M}_s=0.177$~T~\cite{JChen}. 
 For the ultrathin YIG substrate, the Gilbert damping coefficient $\alpha_G\sim 10^{-3}$~\cite{XYWei,HWang} and exchange stiffness $\alpha_{\rm ex}=3\times10^{-16}$~${\rm m}^2$~\cite{JChen}. Besides,  $\mu_0=4\pi\times10^{-7}$ $\rm H/m$ and $\gamma=1.82\times10^{11}$ $\rm s^{-1}\cdot T^{-1}$.

The modulus of effective coupling $|\Gamma({\bf r})|$ under different magnetic configurations $\theta=\{0,\pi,-\pi/4,\pi/4\}$ and $\tilde{\theta}=\{0, 0, -0.056\pi,0.056\pi\}$ are plotted in Table~\ref{2DMSE}(a)-(d),  which show tunable chiralities or non-reciprocities.  In the parallel configuration $\theta=0$, $|\Gamma(\bf r)|$ is symmetric in the $\hat{\bf z}$-direction, but is stronger when $y>0$, implying that the Kittel magnon tends to interact with the substrate magnons propagating to the right. The chirality becomes opposite in the antiparallel configuration $\theta=\pi$. The chirality is altered strongly when $\theta=\pm\pi/4$ as shown in Table~\ref{2DMSE}(c) and (d), where $|\Gamma(\bf r)|$ is asymmetric in both the $\hat{\bf y}$ and $\hat{\bf z}$-direction. These chiralities drive different aggregations of magnonic eigenstates. To show such boundary skin effects, we plot in Table.~\ref{2DMSE}(e)-(h) the spatial distributions of all eigenstates $W({\bf r}_{a,b})=[1/(N_yN_z)]\sum_{\xi}|\psi_{\xi}({\bf r}_{a,b})|^2$. In the collinear parallel and antiparallel configurations, the chirality only drives the skin effect at one edge: as in (e) with $\theta=0$, all the eigenstates pile up at the right edge, but in (f) with $\theta=\pi$, they aggregate at the left edge. While in the non-collinear configuration with $\theta=\pm\pi/4$ as in (g) and (h), all the magnonic eigenstates 
become skewed to the lower-right and upper-right corners, respectively, showing two kinds of non-Hermitian skin effects. These non-Hermitian skin modes are of first order since the inversion symmetry is broken~\cite{Kawabata}, different from the higher-order corner skin modes that need specific symmetry~\cite{KZhang2,Haiping,Kawabata,HYWang}.

\begin{table*}[htp]
\caption{Modulus of the effective coupling $\Gamma({\bf r})$ between Kittel magnons in the nanomagnets [(a)-(d)], corresponding edge or corner aggregations of the magnon eigenstates [(e)-(h)], and spectral windings [(i)-(p)] in different magnetic configurations $\theta=\{0,\pi,\pi/4,-\pi/4\}$. }\label{2DMSE}
	\centering
		\begin{tabular}{cccc}
		\hline
		\toprule
	\makebox[0.08\textwidth][c]{~~~~~Configuration} & \makebox[0.225\textwidth][c]{~~~~~~~~~~~~~~~~~~Coupling constant} &
	 \makebox[0.25\textwidth][c]{~~~~~~~~~Skin effect}  & \makebox[0.28\textwidth][c]{Spectral winding} \\
		\midrule[0.5pt]
		\begin{minipage}[m]{0.09\textwidth}
      \centering
		\thead{$\theta= 0$\\ \\ $({\cal W}_y,{\cal W}_z)=(1,0)$}
		\end{minipage} & \begin{minipage}[m]{0.12\textwidth}			\centering\vspace*{2pt}			\includegraphics[width=3.9cm]{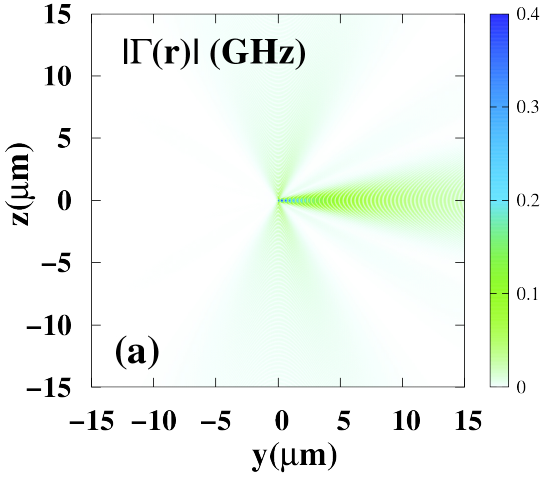}
		\end{minipage} &  \begin{minipage}[m]{0.12\textwidth}
			\centering\vspace*{2pt}
			\includegraphics[width=2.1\textwidth,trim=2cm 0cm 0cm 0.1cm]{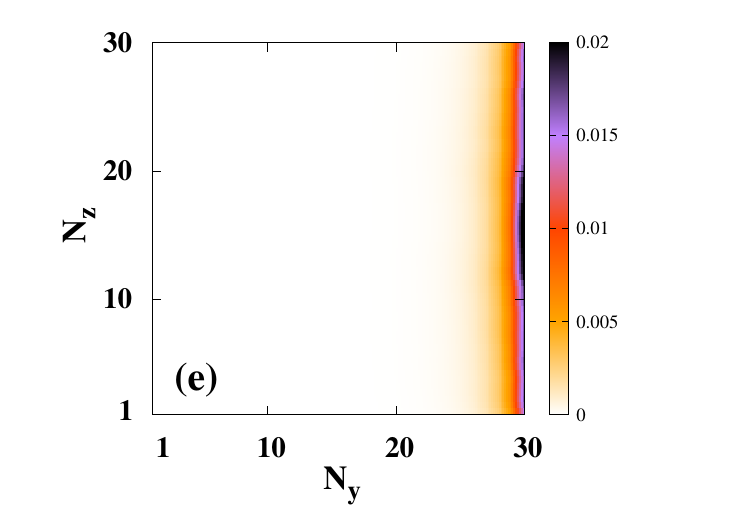}
		\end{minipage} &  \begin{minipage}[m]{0.41\textwidth}
			\centering\vspace*{2pt}
			\hspace*{3pt}
			\includegraphics[width=0.47\textwidth,trim=0cm 0cm 0cm 0.2cm]{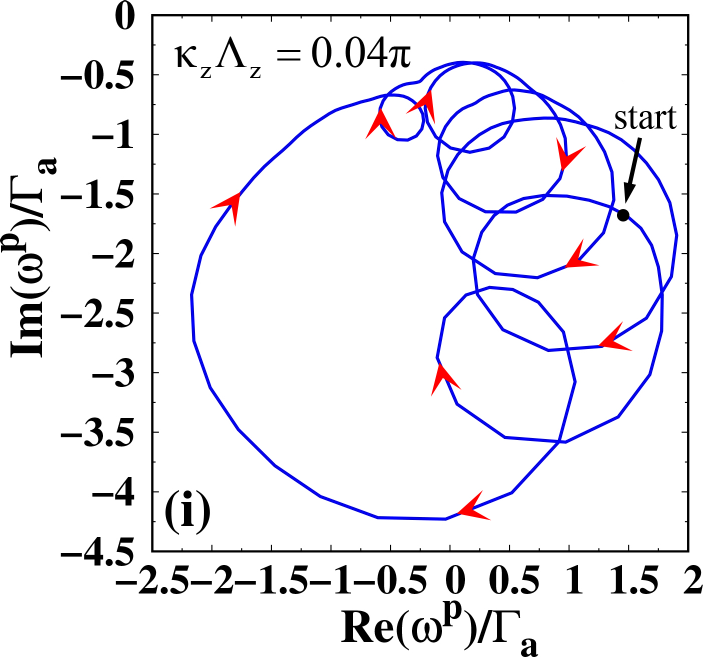}
   \includegraphics[width=0.473\textwidth,trim=0cm 0cm 0cm 0cm]{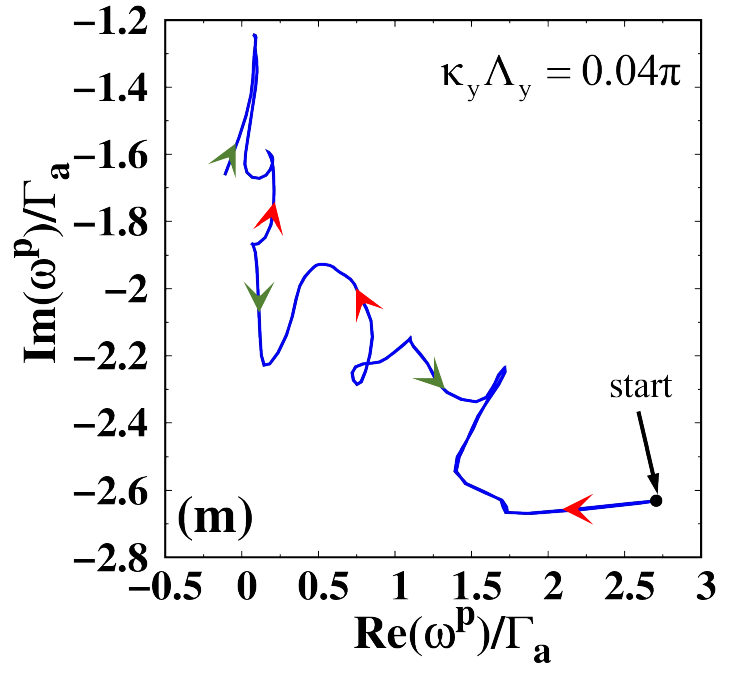}
		\end{minipage} \\
		\midrule[0.5pt]
		\begin{minipage}[m]{0.09\textwidth}
			\centering
	\thead{$\theta= \pi$\\ \\$({\cal W}_y,{\cal W}_z)=(-1,0)$}
		\end{minipage} & \begin{minipage}[m]{0.12\textwidth}
			\centering\vspace*{2pt}
			\includegraphics[width=3.9cm]{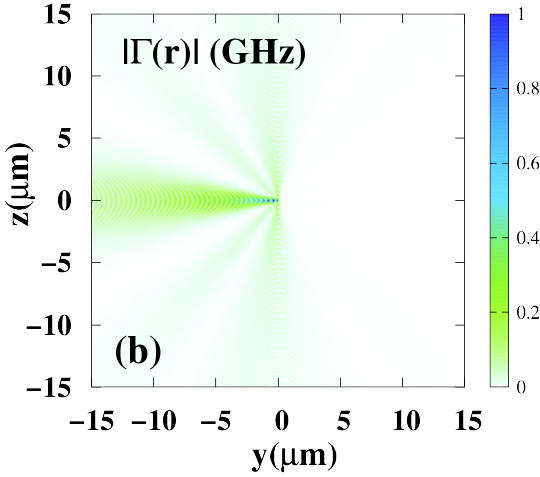}
		\end{minipage} &  \begin{minipage}[m]{0.12\textwidth}
			\centering\vspace*{2pt}
			\includegraphics[width=2.1\textwidth,trim=2cm 0cm 0cm 0.1cm]{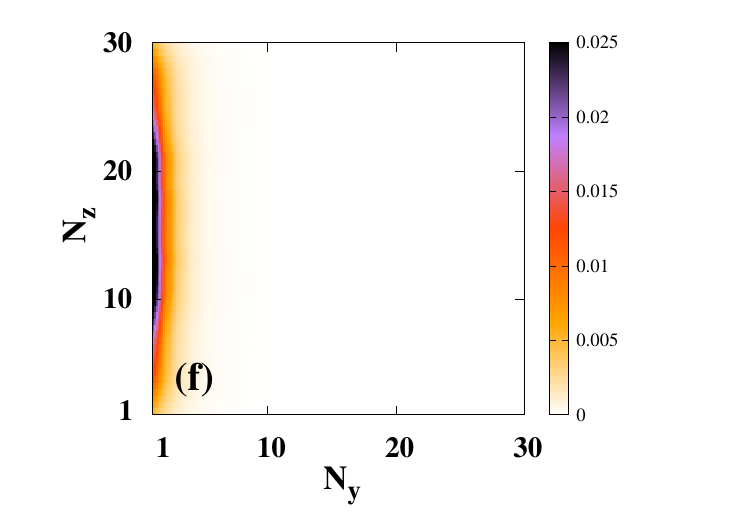}
		\end{minipage} &  \begin{minipage}[m]{0.41\textwidth}
			\centering\vspace*{2pt}
			\hspace*{3pt}
				\includegraphics[width=0.471\textwidth,trim=0cm 0cm 0cm 0.2cm]{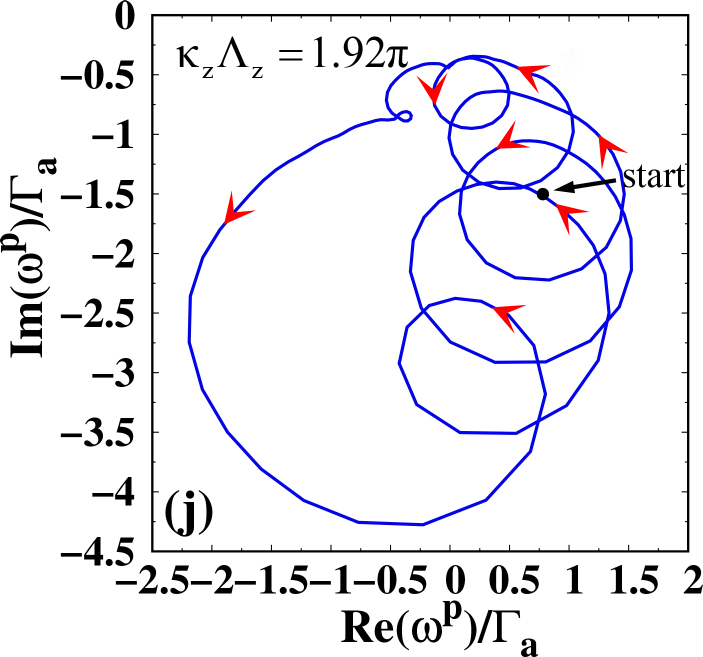}
   \includegraphics[width=0.477\textwidth,trim=0cm 0cm 0cm 0.2cm]{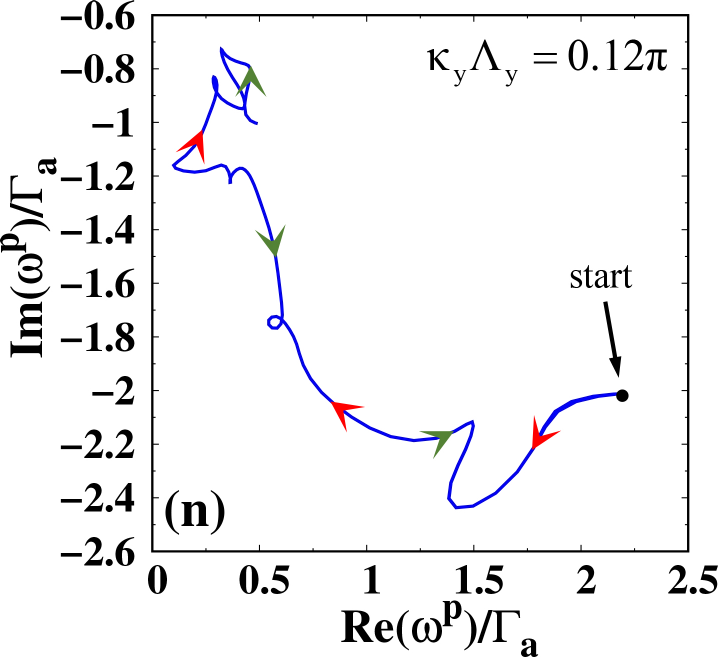}
		\end{minipage} \\
		\midrule[0.5pt]
		\begin{minipage}[m]{0.09\textwidth}
			\centering
		\thead{$\theta= \frac{\pi}{4}$\\ \\ $({\cal W}_y,{\cal W}_z)=(1,-1)$}
		\end{minipage} & \begin{minipage}[m]{0.12\textwidth}
			\centering\vspace*{2pt}
			\includegraphics[width=3.9cm]{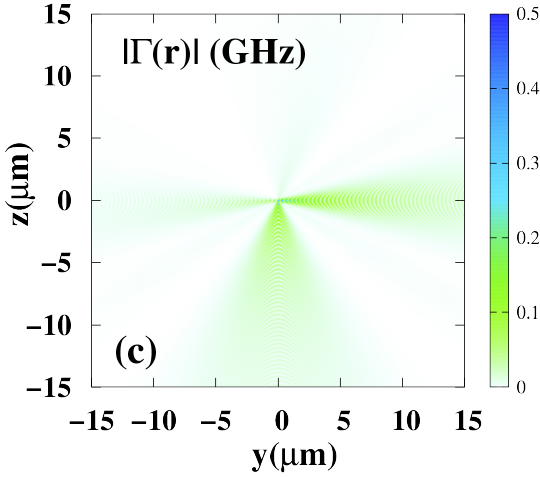}
		\end{minipage} &  \begin{minipage}[m]{0.12\textwidth}
			\centering\vspace*{2pt}
			\includegraphics[width=2.1\textwidth,trim=2cm 0cm 0cm 0.1cm]{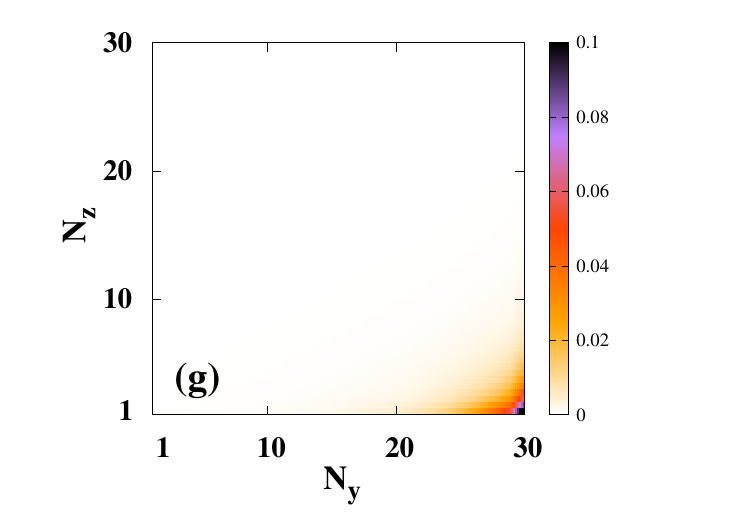}
		\end{minipage} &  \begin{minipage}[m]{0.41\textwidth}
			\centering\vspace*{2pt}
			\hspace*{3pt}
				\includegraphics[width=0.469\textwidth,trim=0cm 0cm 0cm 0.2cm]{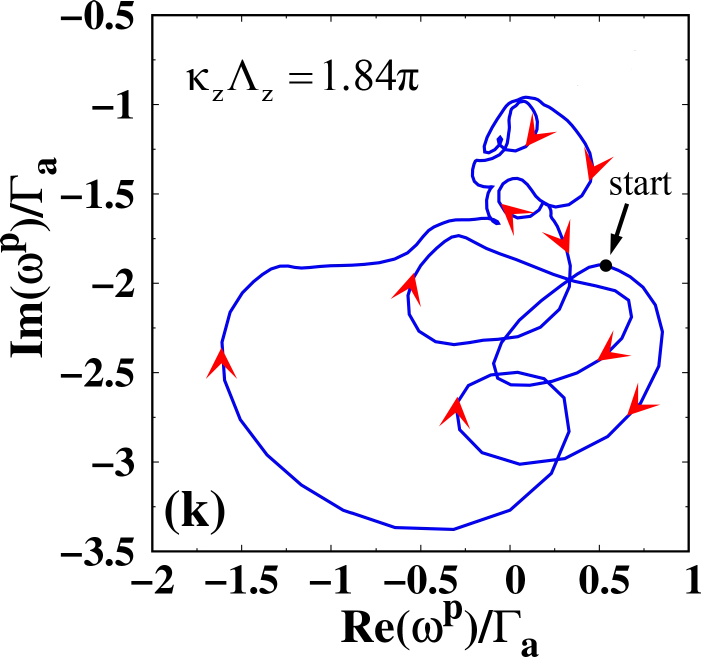}
   \includegraphics[width=0.475\textwidth,trim=0cm 0cm 0cm 0.2cm]{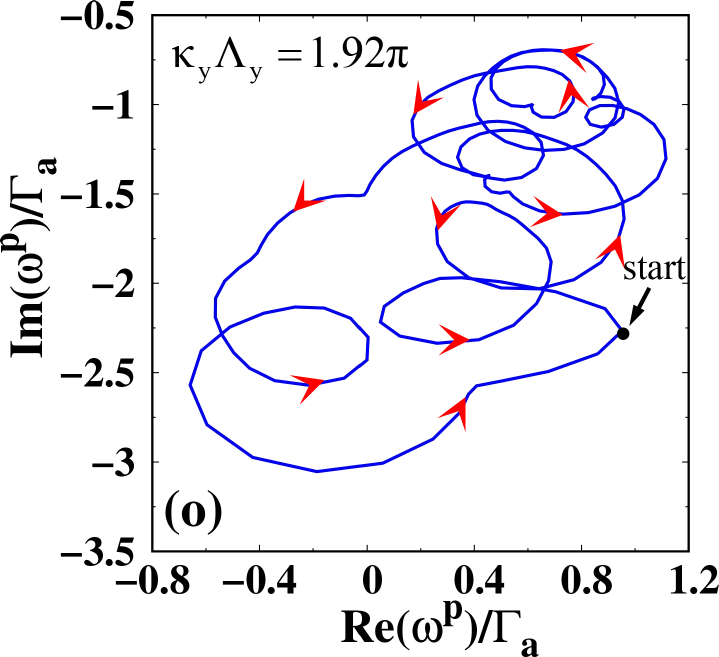}
		\end{minipage} \\
		\midrule[0.5pt]
	\begin{minipage}[m]{0.09\textwidth}
		\centering
		\thead{$\theta= -\frac{\pi}{4}$\\ \\$({\cal W}_y,{\cal W}_z)=(1,1)$}
	\end{minipage} & \begin{minipage}[m]{0.12\textwidth}
		\centering\vspace*{2pt}
		\includegraphics[width=3.9cm]{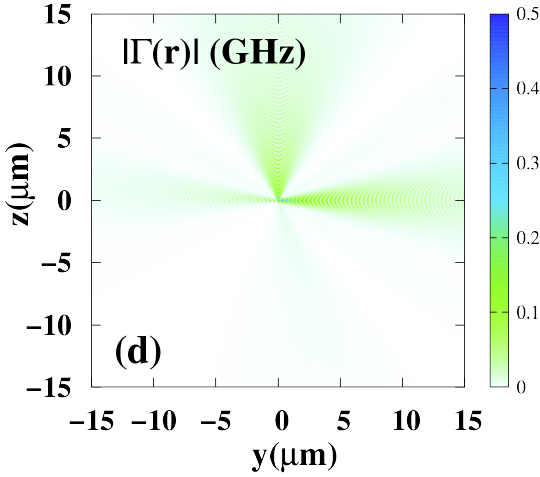}
	\end{minipage} &  \begin{minipage}[m]{0.12\textwidth}
		\centering\vspace*{2pt}
		\includegraphics[width=2.1\textwidth,trim=2cm 0cm 0cm 0.1cm]{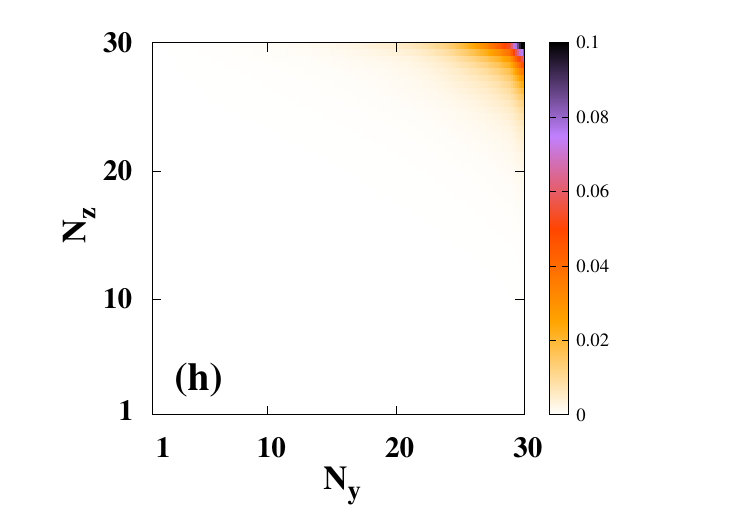}
	\end{minipage} &  \begin{minipage}[m]{0.41\textwidth}
		\centering\vspace*{2pt}
		\hspace*{3pt}
		\includegraphics[width=0.47\textwidth,trim=0cm 0cm 0cm 0.2cm]{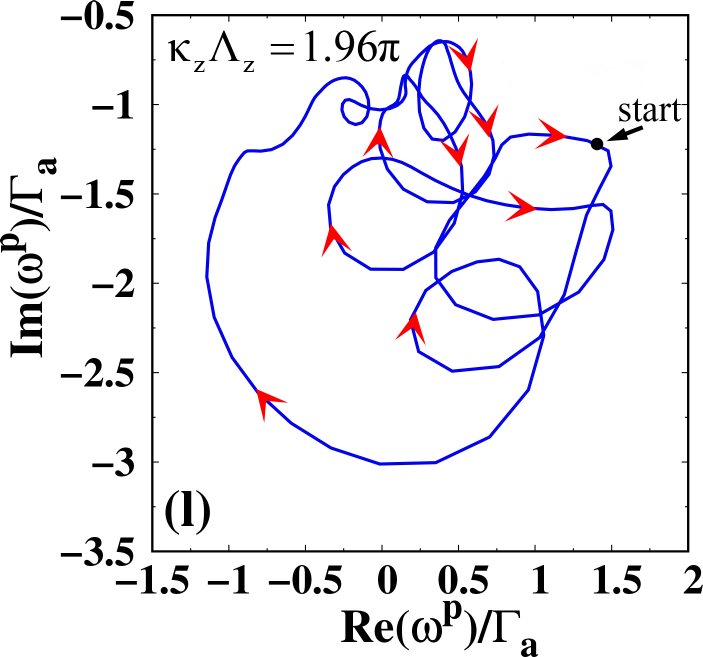}
   \includegraphics[width=0.475\textwidth,trim=0cm 0cm 0cm 0.3cm]{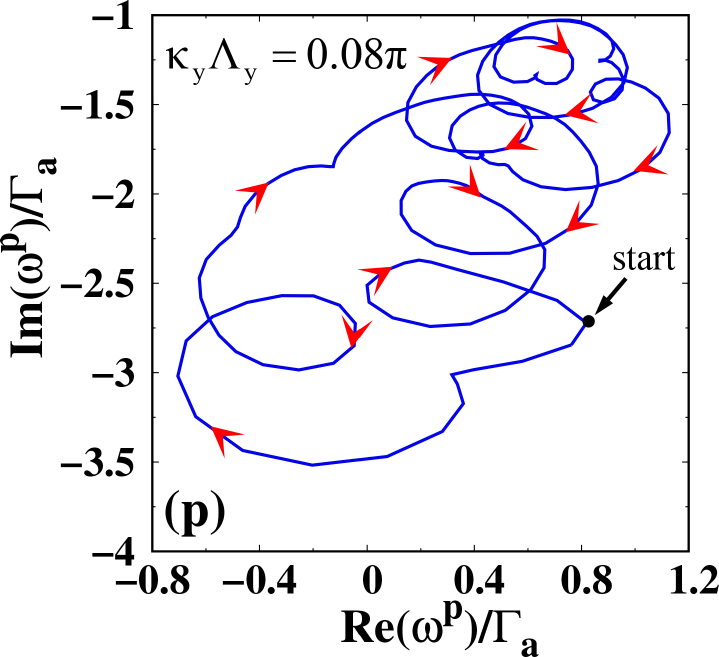}
	\end{minipage} \\
		\toprule
	\end{tabular}

\end{table*}

\textit{Winding tuple and topological characterization.}---As addressed, the winding tuple is a convenient tool to topologically distinguish the edge and corner skin modes in the 2D non-Hermitian skin effect of the non-reciprocal systems~\cite{KZhang2,Haiping,Kawabata,HYWang}. To this end, we address a topological characterization of the aggregations of magnon eigenstates in terms of the winding tuple of the complex frequency spectra under periodic boundary conditions.

\begin{figure}[ht]
\centering
\includegraphics[width=0.49\textwidth]{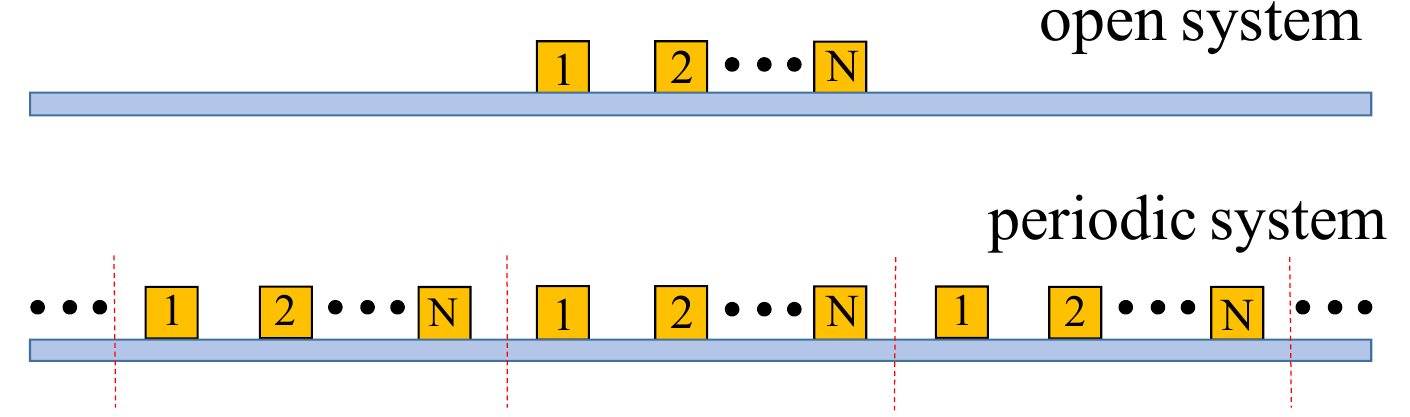}
	\caption{Mapping of the open system with a finite array to the periodic system via repeating the array on the substrate an infinite number of times.}
	\label{periodic_boundary_condition}
\end{figure}

However, before we can turn to this winding tuple, we need to deal with the long-range coupled system rendering the construction of periodic boundary conditions non-trivial since every two magnets couple, differently from the short-range coupled system~\cite{NHatano,Yokomizo,KZhang1}. To solve this issue we propose to map the system with a finite array on the substrate under open boundary conditions to the periodic system by repeating the finite array on the substrate an infinite number of times and requesting the magnon operator in the $a$-th column  and $b$-th row to satisfy periodic condition $\hat{\beta}_{(a,b)}=\hat{\beta}_{(a+N_y,b)}=\hat{\beta}_{(a,b+N_z)}$, as addressed in Fig.~\ref{periodic_boundary_condition} for the one-dimensional situation. Good agreement is obtained in the one-dimensional system, which allows an analytical treatment, where our numerical results agree with the analytical one~\cite{Yu_Zeng}. We refer to the SM~\cite{supplement} for a detailed comparison.

The translational symmetry is recovered when we repeat the block of the nanomagnet array along the $\hat{\bf y}$- and $\hat{\bf z}$-directions indefinitely. We label every block by $\{n_y,n_z\}\in (-\infty,\infty)$ and every nanomagnet in the block by $\{a,b\}$. The magnons in the substrate then interact with the Kittel magnons in all nanomagnets, leading to the Hamiltonian 
\begin{align}
&\hat{H}_p/\hbar=\sum_{n_y,n_z}\sum_{a=1}^{N_y}\sum_{b=1}^{N_z}(\Omega-i\delta_\beta) \hat{\beta}^{(n_y,n_z)\dagger}_{a,b}\hat{\beta}^{(n_y,n_z)}_{a,b}\nonumber\\&+\sum_{\bf k}(\omega_{k}-i\delta_m)\hat{m}^{\dagger}_{\bf k}\hat{m}_{\bf k}+\bigg(\sum_{\bf k}\sum_{n_y,n_z}\sum_{a=1}^{N_y}\sum_{b=1}^{N_z} g_{\bf k}\hat{m}_{\bf k}  \nonumber\\&\times \hat{\beta}^{(n_y,n_z)\dagger}_{a,b}e^{i\big((a+n_yN_y)k_y\Lambda_y+(b+n_zN_z)k_z\Lambda_z\big)}+{\rm H.c.}\bigg),
\label{H_p}
\end{align}
where the phase in the coupling term records the position of the nanomagnet. 
Due to the periodicity, we only need to focus on one block such as the $\{n_y=0,n_z=0\}$ block. Below we denote $\hat{\beta}_{a,b}^{(0,0)}$ by $\hat{\beta}_{a,b}$ for short notation.
By Langevin's equation~\cite{Gardiner,Clerk} and using the effective coupling (\ref{effective_coupling}), we find 
\begin{align}
    &\left(\omega-\Omega+i \delta_\beta\right) \hat{\beta}_{a,b}\nonumber\\
    &=-i\sum_{a',b'}\sum_{n_y,n_z}\Gamma({\bf r}_{a,b}-{\bf r}_{a'+n_yN_y,b'+n_zN_z})\hat{\beta}_{a',b'}\nonumber\\
    &=-i\sum_{a',b'}\Gamma^p({\bf r}_{a,b}-{\bf r}_{a',b})\hat{\beta}_{a',b'},
    \label{beta_p2}
\end{align}
where ${\bf r}_{a,b}=a \Lambda_y\hat{\bf y}+b \Lambda_z\hat{\bf z}$ is the position of the $(a,b)$-th nanomagnet and in the second line we impose the periodic condition $\beta_{a,b}^{(n_y,n_z)}=\beta_{a,b}^{(0,0)}$. 
 \[\Gamma^p({\bf r}_{a,b}-{\bf r}_{a',b'})=\sum_{n_y,n_z}\Gamma({\bf r}_{a,b}-{\bf r}_{a'+n_yN_y,b'+n_zN_z})
 \]
 is periodic in both the $\hat{\bf y}$- and $\hat{\bf z}$-directions since $\Gamma^p({\bf r})=\Gamma^p({\bf r}+N_y\Lambda_y\hat{\bf y})=\Gamma^p({\bf r}+N_z\Lambda_z\hat{\bf z})$.

We then find from Eq.~(\ref{beta_p2}) the elements of the Hamiltonian matrix of the periodic system, which under the on-shell approximation $\omega\rightarrow\Omega$ read
\begin{align}
	& {\cal H}_{\rm eff}^p\big|_{a=a',b=b'}=\Omega-i \delta_\beta-i\Gamma^p({\bf r}=0),\nonumber\\
	& {\cal H}_{\rm eff}^p\big|_{a\ne a'~{\rm or}~b\neq b'}=-i\Gamma^p({\bf r}_{a,b}-{\bf r}_{a',b'}).
	\label{Heff^p}
\end{align}
Due to the periodicity of $\Gamma^p({\bf r})$ the  eigenfunctions of matrix ${\cal H}^p_{\rm eff}$ are the plane waves
\begin{align}
&{\pmb\psi}^p_{\kappa_y,\kappa_z}=\frac{1}{\sqrt{N_yN_z}}\left(e^{i(\kappa_y\Lambda_y+\kappa_z\Lambda_z)},e^{i(\kappa_y\Lambda_y+2\kappa_z\Lambda_z)},\cdots,\right.\nonumber\\&\left.e^{i(\kappa_y\Lambda_y+N_z\kappa_z\Lambda_z)},e^{i(2\kappa_y\Lambda_y+\kappa_z\Lambda_z)},\cdots,e^{i(N_y\kappa_y\Lambda_y+N_z\kappa_z\Lambda_z)}\right)^T,
\end{align}
where $\kappa_y\equiv 2\pi l_y/(N_y\Lambda_y)$ and $\kappa_z\equiv 2\pi l_z/(N_z\Lambda_z)$ are real with integers $l_y=\{1,2,...,N_y\}$ and $l_z=\{1,2,...,N_z\}$.
It obeys ${\cal H}^p_{\rm eff}{\pmb\psi}^p_{\kappa_y,\kappa_z}=\omega^p({\kappa_y,\kappa_z}){\pmb\psi}^p_{\kappa_y,\kappa_z}$, where the eigenfrequency 
\begin{align}
{\omega}^p({\kappa_y,\kappa_z})&=\Omega-i\delta_\beta-i\sum_{a=0}^{N_y-1}\sum_{b=0}^{N_z-1}\Gamma^p(-a \Lambda_y\hat{\bf y}-b \Lambda_z\hat{\bf z})\nonumber\\&\times e^{i(a\kappa_y\Lambda_y+b\kappa_z\Lambda_z)}.
\end{align}

Since the complex spectra ${\omega}^p({\kappa_y,\kappa_z})$ are  functions of two
real wave numbers $\kappa_y$
and $\kappa_z$, they have a complicated distribution on the complex plane. The conventional spectra topology with the winding number in the one-dimensional system is still convenient to characterize the topological origin of the skin effect. Here we use it to
 characterize the 2D non-Hermitian skin effect by fixing one component of $(\kappa_y\Lambda_y,\kappa_z\Lambda_z)$ at any (convenient) value and monitor the evolution of $\omega^p(\kappa_y,\kappa_z)$ on the complex plane when the other wave number evolves by a period.  Accordingly, we define the topological winding tuple $({\cal W}_y,{\cal W}_z)$ by fixing, respectively, $\kappa_z\Lambda_z$ and $\kappa_y\Lambda_y$ for the entries of the tuple:
 \begin{align}
{\cal W}_{i=\{y,z\}}=\left\{\begin{array}{cc}
0,\quad \quad \quad \quad \quad~~ {\rm if}~{\forall} \omega_0,~Q_i=0 \\
-Q_i/|Q_i|,\quad \quad {\rm if}~{\exists}\omega_0,~Q_i\ne 0
\end{array},\right.
\label{winding}
\end{align}
where with respect to the reference frequency $\omega_0$
\[Q_{i=\{y,z\}}=\int_{0}^{2\pi}\frac{d}{d(\kappa_i\Lambda_i)} \arg[\omega^p(\kappa_y,\kappa_z)-\omega_0]d(\kappa_i\Lambda_i).\]
 When the spectra do not form a loop, ${\cal W}_i=0$; otherwise ${\cal W}_i=1 (-1)$ for the clockwise (anticlockwise) evolution of the frequency spectra,  which can be computed by properly choosing $\omega_0$ on the complex plane.

The winding tuple $({\cal W}_{y},{\cal W}_{z})$ precisely characterizes different edge or corner localization in the 2D non-Hermitian skin effect of the non-reciprocal or chiral systems. When both two indexes vanish, no 2D non-Hermitian skin effect occurs; when only one of them is nonzero, the magnon eigenmodes are localized on one of the edges, i.e. upper, lower, left, and right skin modes that are characterized, respectively, by $\{{\cal W}_y,{\cal W}_z\}=\{0,1\},\{0,-1\},\{-1,0\}$, and $\{1,0\}$; when both exist, the skin modes pile up at one of the corners, with the upper-left, lower-left, upper-right, lower-right corner modes characterized, respectively, by $\{{\cal W}_y,{\cal W}_z\}=\{-1,1\},\{-1,-1\},\{1,1\}$, and $\{1,-1\}$. 
 This is justified by the numerical calculation in Table.~\ref{2DMSE}(i)-(p)  with $N_y=N_z=250$ for the spectra winding when fixing one of $\kappa_y$ and $\kappa_z$. For the edge skin effect when $\theta=\{0,\pi\}$ one component of the winding numbers vanishes; while for the corner skin effect when $\theta=\pm \pi/4$, both winding numbers are nonzero that governs the position that the magnonic eigenstates localize.

\textit{Discussion.}---In conclusion, we predict the edge or corner skin effects of magnons in the nanomagnetic array that act as magnetic dipoles on a high-quality magnetic insulating substrate and fully characterize their topological origin in terms of winding tuples. Such an approach can be extended to the three-dimensional case with a winding three-tuple and so on for a long-range coupled system of regular shape. The insights obtained in magnonics, where magnetic dipoles are exploited, should straightforwardly apply to analogous electric dipoles that are coupled in a long-range way, for instance in chiral photonics~\cite{Lodahl,ZLan,Ozawa,LingLu} or plasmonics~\cite{Rodriguez,Petersen}.%, cold atoms, and phononics. 

\begin{acknowledgments}
This work is financially supported by the National Natural Science Foundation of China under Grant No. 12374109, and the startup grant of Huazhong University of Science and Technology (Grants  No.~3004012185 and 3004012198). DMK acknowledges  funding by the DFG under RTG 1995, within the Priority Program SPP 2244 ``2DMP'' --- 443273985 and under Germany's Excellence Strategy - Cluster of
Excellence Matter and Light for Quantum Computing (ML4Q) EXC 2004/1 -
390534769. 
\end{acknowledgments}

\end{document}